\documentclass[12pt, prd, showpacs]{revtex4}
\usepackage{amssymb}
\usepackage{amsmath}

\input{tcilatex}

\begin{document}

\title{Distorted vacuum black holes in the canonical ensemble}
\author{O. B. Zaslavskii}
\affiliation{Department of Physics and Technology, Kharkov V.N. Karazin National
University, 4 Svoboda Square, Kharkov 61022, Ukraine}
\affiliation{Institute of Mathematics and Mechanics, Kazan Federal University, 18
Kremlyovskaya St., Kazan 420008, Russia}
\email{zaslav@ukr.net }

\begin{abstract}
We consider a vacuum static spacetime in a finite size cavity. On the
boundary, we specify a metric and a finite constant local temperature $T$.
No spherical or any other spatial symmetry is assumed. We show that (i)
inside a cavity, only a black hole or flat spacetime are possible, whereas a
curved horizonless regular space-time is excluded, (ii) in the limit when
the horizon area shrinks, the Hawking temperature diverges, (iii) for the
existence of a black hole, $T$ should be high enough. When $T\rightarrow
\infty $, a black hole phase is favorable thermodynamically. Our
consideration essentially uses the coordinate system introduced by Israel in
his famous proof of the uniqueness theorem.
\end{abstract}

\maketitle

\section{Introduction}

For consistent thermodynamics of self-gravitating systems, a finite size is
a key point to be taken into account. This was demonstrated first by York,
Jr. \cite{york86} who showed that account for the boundary resolves the
problems that seem to prevent the construction of the canonical ensemble for
Schwarzschild black holes and leads to new interesting features. In
particular, it turned out that for a given temperature $T$ on the boundary
and the radius $R$ of the cavity, there exist two branches of black hole
metrics, if $T$ is high enough, one of two branches being locally or even
globally stable. This also showed that in discussing thermal nucleation of
black holes from hot empty space-time \cite{gross}, one cannot neglect the
presence of the boundary without which the thermal ensemble does not exist
at all.

Further, some general formulas describing thermal properties of the
gravitating ensembles were derived \cite{marty} - \cite{6}. Unfortunately,
application of the general formalism in concrete analysis was done for
spherically symmetric metrics only \cite{brad}, \cite{jose} since the
explicit expression for distorted black holes is, as a rule, absent or too
complicated. This somewhat slowed down further progress in this area.

The aim of the present note is to show that there are some simple but useful
properties of distorted black holes in the canonical ensemble that can be
inferred with minimum information, almost "from nothing". It turned out that
this step can be performed due to using a simple and elegant coordinate
frame used by W. Israel \cite{isr} in proving his famous uniqueness theorems
for black holes. Now, however, one should bear in mind that in contrast to 
\cite{isr}, our space-time is not asymptotically flat due to the presence of
the boundary and this is the crucial point.

\section{Basic equations}

In the coordinate frame of Ref. \cite{isr}, the metric can be written in the
form%
\begin{equation}
ds^{2}=-V^{2}dt^{2}+\rho ^{2}dV^{2}+\gamma _{ab}dx^{a}dx^{b}\text{.}
\label{met}
\end{equation}%
Here, $a,b=1,2$. It is supposed that the metric coefficients do not depend
on $t$. It is seen from (\ref{met}) that 
\begin{equation}
\rho ^{-2}=(\nabla V)^{2}\text{.}
\end{equation}%
In the horizon limit, 
\begin{equation}
\rho \rightarrow \rho _{H}=\frac{1}{\kappa }\text{,}  \label{kappa}
\end{equation}%
where $\kappa $ is the surface gravity (see, e.g.. eqs. 105 - 107 in \cite%
{vis}).

Then, the Hawking temperature%
\begin{equation}
T_{H}=\frac{\kappa }{2\pi }=\frac{1}{2\pi \rho _{H}}\text{.}  \label{surf}
\end{equation}%
Eq. (32) of \cite{isr} gives us%
\begin{equation}
\frac{\partial }{\partial V}(\frac{\sqrt{\gamma }}{\rho })=0\text{,}
\label{gv}
\end{equation}%
whence%
\begin{equation}
\rho =\sqrt{\gamma }C(x^{1},x^{2})\text{.}  \label{roc}
\end{equation}

Eq. (39) of \cite{isr} tells us that the Kretschmann scalar $Kr$ in the
vacuum space-time

\begin{equation}
\frac{Kr}{8}=\frac{1}{8}R_{ABCD}R^{ABCD}=(V\rho ^{)^{-2}}[K_{ab}K^{ab}+2\rho
^{-2}\rho _{;a;b}\rho ^{a;b}+\rho ^{-4}\left( \frac{\partial \rho }{\partial
V}\right) ^{2}]\text{.}  \label{kr}
\end{equation}

\section{Properties of the canonical ensemble}

Usually, when dealing with thermodynamic description of some space-time, one
first finds its metric (or takes the already known one) and only afterwards
ascribes thermodynamic parameters to the system. Meanwhile, a coherent
approach to finite size thermodynamic implies something quite different.
What is done in the problem are boundary data that are specified on some
surface that is not necessarily spherical. These includes the boundary
metric $\gamma _{ab}$ and the local temperature. Also, \ for specifying a
solution of field equations, we need $\rho $ and $K_{ab}$. Integrating
equations of motion in the inward direction, one can in principle (but not
in practice) find these solutions inside a cavity. Meanwhile, for our
purpose we need much less information (see below) that require only one
equation plus regularity conditions.

Statement 1. If inside a cavity the vacuum space-time is not flat, it cannot
be horizonless.

Proof. If the space-time is flat, $V\equiv 1$ and cannot be taken as an
independent variable. We assume that it is not flat, so the metric in the
form (\ref{met}) can be used. Let us suppose that there is no horizon, so
the metric has a regular centre. This means that for some $V=V_{1}>0$ the
quantity $\sqrt{\gamma }=0$. Then, according to (\ref{roc}), $\rho
\rightarrow 0$ as well when $V\rightarrow V_{1}$. In (\ref{kr}) this entails
that the finiteness of $Kr$ requires%
\begin{equation}
\frac{\partial \rho }{\partial V}\sim \rho ^{3}\text{, }\frac{1}{\rho ^{2}}%
\sim V+const\text{, }\rho (V_{1})\neq 0\text{.}
\end{equation}%
This is in contradiction with (\ref{roc}).

Thus there are only two phases take part in thermodynamic competition: the
flat space-time and a black hole.

Statement 2. A vacuum black hole \ with a finite horizon area cannot be
extremal ($\kappa \neq 0$).

Proof. If $\kappa =0$, it follows from (\ref{kappa}) that $\rho _{H}=\infty $%
. Then, it follows from (\ref{roc}) that $\gamma _{H}=\infty $ as well in
contradiction with the assumption.

Statement 3. If the horizon area shrinks, the Hawking temperature in this
limit grows unbounded: 
\begin{equation}
\lim_{A_{H}\rightarrow 0}T_{H}=\infty \text{.}  \label{th}
\end{equation}

Proof. The area $A$ of any equipotential surface $V=V_{1}=const$ is equal to%
\begin{equation}
A=\int dx^{1}dx^{2}\sqrt{\gamma }\text{.}
\end{equation}

Here, $\gamma =\gamma (V,x^{1},x^{2})$, integration is performed in some 
\textit{fixed} intervals $a_{1}\leq x^{1}\leq b_{1}$, $a_{2}\leq x^{2}\leq
b_{2}$ (for instance, one can choose the analogue of angular variables).
Therefore, the condition that $A\rightarrow 0$ entails that $\sqrt{\gamma }%
\rightarrow 0$ as well. Then, it follows from (\ref{roc}) that $\rho
_{H}\rightarrow 0$, so that $T_{H}\rightarrow \infty $ according to (\ref%
{surf}). Eq. (\ref{th}) generalizes the corresponding property of the
Schwarzschild metric, where $T_{H}=(4\pi r_{+})^{-1}$, $r_{+}$ being the
horizon radius.

It is essential that we deal with equipotential surfaces. For comparison let
us consider, say, the Schwarzschild metric. We can take an arbitrary point
and encircle it by a small sphere with minimum and maximum value of the
standard Schwarzschild coordinate $r_{1}$ and $r_{2}$. Obviously, such a
sphere is not equipotential surface and the above reasonings do not apply.
When $r_{2}\rightarrow r_{1}$, the area vanishes although $\sqrt{\gamma }%
=r^{2}\sin ^{2}\theta $ remains separated from zero.

Statement 4. Black hole solutions are possible only in the high temperature
phase, $T>T_{m}$, where the concrete value $T_{m}$ is determined by the
boundary conditions (and thus cannot be found in a general form).

Proof. Let $\beta \equiv T^{-1}$ be the inverse temperature on the boundary.
It follows from the Tolman formula that%
\begin{equation}
\beta =\beta _{0}V_{B}\text{,}  \label{bv}
\end{equation}%
where $\beta _{0}=T_{H}^{-1}$ is a constant, $V_{B}$ is the value of $V$ on
the boundary.

For a black hole, the horizon area $A_{H}$ lies in the interval $%
0<A_{H}<A_{B}$, where $A_{B}$ corresponds to the boundary. It follows from (%
\ref{th}) that in the limit when the horizon shrinks to the point, so $%
A_{H}\rightarrow 0$, the quantity $\beta \rightarrow 0$ due to the factor $%
\beta _{0}$. On the other hand, if a black hole occupies almost the whole
cavity, the boundary almost coincides with the horizon, so $V_{B}$
approaches $V$ on the horizon where it is equal to zero. Now, $\beta
\rightarrow 0$ due to the second factor in (\ref{bv}). Thus the quantity $%
\beta $ vanishes in both limits: for the minimum possible area (equal to
zero) and the maximum one (corresponding to the boundary). Therefore, in
some intermediate point $A_{H}=A_{m}$ the inverse temperature $\beta $
should pass through the maximum point equal to some $\beta _{m}=\beta
(A_{m}) $. This proves the statement. For extremal black holes it would be
possible for $\beta \,\ $to diverge but we deal now with nonextremal ones.

Statement 5. For a given $T$, there exist at least two branches of black
holes, one of which (at least in some interval of temperatures) is locally
stable.

Proof. As $\beta (A)$ has two zeros at $A_{H}=0$ and $A_{H}=A_{B}$, the
branch $A_{m}<A_{H}<A_{B}$ has $\frac{d\beta }{dA_{H}}<0$, if it is
monotonic. Then, $\frac{dT}{dA_{H}\text{,}}>0$. Meanwhile, the heat capacity 
$C=\frac{dE}{dT}=T\frac{dS}{dT}=\frac{T}{4}\frac{dA_{H}}{dT}$, where $E$ is
the energy, $S=\frac{A_{H}}{4}$ is the Bekenstein-Hawking entropy, and the
first law $dE=TdS$ was used. We see that $C>0$, so the solution is locally
stable. Even if $\beta $ as a function of $A_{H}$ is not monotonic, the part
of $\beta (A)$ in the vicinity of $A_{B}$ is monotonically decreasing, and
the above reasonings apply.

Statement 6. For sufficiently high temperature, a black hole phase is
favorable both locally and globally.

Proof. It is quite obvious that for very high temperatures configuration
with a black hole will dominate. Indeed, in the Eucldiean action approach,
the free energy $F=TI$, the Euclidean action for a black hole topology $%
I=\beta E-S$ \cite{can}. When $\beta \rightarrow 0$, the first term in $I$
is negligible, so $I<0$. Meanwhile, for a hot empty space $I=0\,.$ As a
black hole exists for sufficiently high temperature $T>T_{m}$ and becomes
thermodynamically favorable in the limit $T\rightarrow \infty $, there
exists some $T_{1}>T_{m}$ such that for $T>T_{1}$ a black hole not only
stable locally (see above) but also globally. Statements 4 - 6 generalize
the corresponding properties for Schwarzschild black holes \cite{york86}.

\section{Conclusions}

Thus we showed that, defining boundary data, the canonical ensemble does
exist if a temperature is high enough. Moreover, inside a cavity a black
hole is favorable thermodynamically for sufficiently big temperature. This
is done in a quite general approach without appealing to the explicit form
of solutions that, generically, cannot be found at all. In doing so, the
full system of Einstein equations was not used directly. We relied only on
one relation (\ref{gv}), and the expression for the Kretschmann scalar (\ref%
{kr}) for vacuum space-times (in which, though, the validity of Einstein
equations was already taken into account indirectly). We also proved
rigorously that for vacuum black holes the Hawking temperature diverges when
the horizon area shrinks. It would be of interest to check discussed
properties for nonvacuum backgrounds and apply them to nucleation of black
holes in cosmological problems.

\begin{acknowledgments}
This work was funded by the subsidy allocated to Kazan Federal University
for the state assignment in the sphere of scientific activities. O. Z. also
thanks for support SFFR, Ukraine, Project No. 32367.
\end{acknowledgments}

\end{document}